\documentclass{article}

\usepackage{arxiv}

\usepackage[utf8]{inputenc} 
\usepackage[T1]{fontenc}    
\usepackage[hidelinks]{hyperref}       
\usepackage{url}            
\usepackage{booktabs}       
\usepackage{amsfonts}       
\usepackage{nicefrac}       
\usepackage{microtype}      
\usepackage{lipsum}
\usepackage{graphicx}
\usepackage{subcaption}
\graphicspath{ {./images/} }

\hypersetup{
  colorlinks   = true, 
  urlcolor     = blue, 
  linkcolor    = blue, 
  citecolor   = blue 
}

\usepackage[square,numbers]{natbib}

\usepackage[dvipsnames]{xcolor}

\usepackage{tikz}
\NewDocumentCommand{\rot}{O{90} O{1em} m}{\makebox[#2][l]{\rotatebox{#1}{#3}}}%
\def\halfcheckmark{\tikz\draw[scale=0.25,fill=black](0,.35) -- (.25,0) -- (1,.7) -- (.25,.15) -- cycle (0.75,0.2) -- (0.77,0.2)  -- (0.6,0.7) -- cycle;}

\usepackage{titling}
\predate{}
\postdate{}

\title{Trackintel: An open-source Python library for human mobility analysis}
\author{Henry Martin${}^{1,2,*}$, Ye Hong${}^{1, *}$, Nina Wiedemann${}^{1, *}$, Dominik Bucher${}^{3, **}$, Martin Raubal${}^{1}$}

\date{}

\begin{document}
\maketitle

${}^1$ Institute of Cartography and Geoinformation, ETH Zurich, Zurich, Switzerland \\
${}^2$ Institute of Advanced Research in Artificial Intelligence (IARAI), Vienna \\
${}^3$ c.technology, Tessinerplatz 7, 8002 Zurich, Switzerland \\
${}^*$ Authors contributed equally. Order was determined randomly \\
${}^{**}$ The majority of this work was done while the author was at the Chair of Geoinformation Engineering, ETH Zurich. \\

\begin{abstract}
Over the past decade, scientific studies have used the growing availability of large tracking datasets to enhance our understanding of human mobility behavior. 
However, so far data processing pipelines for the varying data collection methods are not standardized and consequently limit the reproducibility, comparability, and transferability of methods and results in quantitative human mobility analysis.
This paper presents Trackintel, an open-source Python library for human mobility analysis. Trackintel is built on a standard data model for human mobility used in transport planning that is compatible with different types of tracking data.
We introduce the main functionalities of the library that covers the full life-cycle of human mobility analysis, including  processing steps according to the conceptual data model, read and write interfaces, as well as analysis functions (e.g., data quality assessment, travel mode prediction, and location labeling).
We showcase the effectiveness of the Trackintel library through a case study with four different tracking datasets.
Trackintel can serve as an essential tool to standardize mobility data analysis and increase the transparency and comparability of novel research on human mobility.
\end{abstract}



\section{Introduction}
Human mobility studies using large-scale human digital traces have boomed over the last decade.
On the collective level, researchers revealed that human movement can be universally described using statistical distributions, i.e., the power-law distribution of consecutive displacements~\citep{brockmann_scaling_2006, rhee_levy-walk_2011}, stationary time between displacements~\citep{rhee_levy-walk_2011, song_modelling_2010}, and characteristic distance travelled by individuals (i.e., the radius of gyration)~\citep{gonzalez_understanding_2008, pappalardo_returners_2015}. Moreover, it has been shown that individuals exhibit markedly regular location visitation patterns~\citep{schneider_unravelling_2013} with high theoretical predictability~\citep{song_limits_2010}. People spend most of their time in a few locations~\citep{gonzalez_understanding_2008, song_modelling_2010} and maintain a stable number of important locations over time~\citep{alessandretti_evidence_2018}. 

To a large extent, this progress can be attributed to the widespread availability of large mobility datasets stemming from information and communications technology (ICT) and location based services (LBS) that are now integrated in many aspects of our daily life~\citep{huang_location_2018, kesler_geoprivacy_2018}.
Aside of the progress on the analysis of human movement itself, the increased availability of tracking data has led to a rapid growth of studies that use human mobility data to study phenomena related to human mobility, such as understanding of residential income segregation~\citep{moro_mobility_2021}, quantifying urbanization levels and city livability~\citep{bassolas_hierarchical_2019}, urban sensing~\cite{ahas2015everyday}, developing infrastructure for sustainable mobility~\cite{xu_planning_2018} and responding to epidemic spreading~\citep{chang_mobility_2021}.
However, the \textit{raw} digital traces are often not the targeted unit of analysis; for example, a location where people perform an activity can not directly be derived from GPS track points or mobile phone tower data. Studies thus employ various steps to preprocess data into the desired format. These steps and their outcome are often different across studies~\citep{chen2016promises} due to the variety of the datasets and the different understanding of the definitions, which has led to a vast collection of dataset-specific preprocessing and analysis methods.
%
For example, the study by~\citet{feng_deepmove_2018}, which proposes the DeepMove model that is now widely accepted as a deep learning baseline model for next location prediction~\citep{luca_survey_2021}, generally regards each raw position record as a \textit{location} and does not perform preprocessing. However, focusing on the same problem, \citet{urner_assessing_2018} extract \textit{staypoints} (i.e., all the points where a user stayed for at least a certain duration) from GPS track points and further aggregate them into locations using the \textit{k}-means algorithm. \citet{solomon_analyzing_2021} apply a similar processing concept but introduce the mean shift algorithm to detect staypoints, which are then merged into locations according to a distance threshold.
These examples show how the lack of a standard movement model definition and a common preprocessing standard limit the reproducibility and comparability of the methods and analysis results.

To address these problems, we present Trackintel, an open-source python library for the processing and analysis of movement data. Trackintel is based on an established model for human mobility taken from transport planning, and implements the most important preprocessing steps. It further provides various analysis, visualization and support functions. Due to the versatility of the data model, Trackintel standardizes preprocessing steps for many types of tracking data that characterize the movement of individuals. It thereby greatly simplifies quantitative research based on tracking data and increases its reproducibility.
The remainder of the paper is structured as follows.  Section~\ref{sec:related_work} provides an overview of existing libraries for the analysis and the preprocessing of movement data. Section~\ref{sec:trackintel_framework} first introduces the hierarchical model for human mobility analysis and describes its implementation in Trackintel. This section then proceeds to present the most important functionalities of Trackintel to process movement data. In section~\ref{sec:case_study} we showcase the capabilities of Trackintel to simplify the analysis and comparison of several different tracking datasets. Finally, we summarize and conclude this work in section~\ref{sec:conclusion}.

\section{Related work: Libraries for movement data} \label{sec:related_work}


\begin{table}[h!]
\centering
\caption[Comparison of movement data libraries.]{Comparison of movement data libraries. Packages are predominantly available open source in R and Python and they are compared with regards to their focus, documentation and functionality. While other movement analysis libraries already provide well-maintained and documented code with rich functionality for trajectory analysis, only Trackintel provides robust and flexible methods to aggregate trajectories into locations, trips and tours. \\
(\checkmark / \halfcheckmark / x : available / partially available / not available)}
\resizebox{\textwidth}{!}{

\begin{tabular}{l|lclllllllllllll}
\toprule
                                      \textbf{Package name} &                                        \textbf{Focus} &\rot{\begin{tabular}[x]{@{}c@{}}Programming language \\ Python (P) \end{tabular}}
                                      & \rot{Documentation score} & 
                                      \rot{\begin{tabular}[x]{@{}c@{}}Test coverage (* / **: not \\  reported but low / high) \end{tabular}}

                                      & \rot{individual (I) / collective (C)} & \rot{\begin{tabular}[x]{@{}c@{}}human (H), animal (A)\\and/or object(O) \end{tabular}} & \rot{Staypoint detection} & \rot{Aggregation to location} & \rot{Aggregation to trips} & \rot{Aggregation to tours} & \rot{Tracking quality assessment} & \rot{Transport mode labelling} & \rot{Home and work labelling} & \rot{Visualization} &
                                      \rot{\begin{tabular}[x]{@{}c@{}}Trajectory statistics \\ (- / + / ++ : none / basic / rich)
                                      \end{tabular}}
                                      \\
\midrule
                                        Trackintel & Human mobility analysis &                          P &                         6 &                98\% &                                     I &                                         H &                \checkmark &                    \checkmark &                 \checkmark &                 \checkmark &                        \checkmark &                     \checkmark &                    \checkmark &          \checkmark &                          +  \\
\midrule
Scikit-mobility~\cite{pappalardo2019scikitmobility} &                           Human mobility analysis  &                          P &                         5 &                  ** &                                     I &                                         H &                \checkmark &                    \checkmark &                          x &                          x &                    \halfcheckmark &                              x &                \halfcheckmark &          \checkmark &                         ++  \\
\midrule
      Movingpandas~\cite{graser_movingpandas_2019} &                             Movement data analysis &                          P &                         6 &                96\% &                                     I &                                       H/A/O &                \checkmark &                             x &             \halfcheckmark &                          x &                                 x &                              x &                             x &          \checkmark &                         ++  \\
\midrule
              PyMove~\cite{sanches2019arquitetura} &  \begin{tabular}[x]{@{}l@{}}Querying and\\visualizing trajectories \end{tabular} &                          P &                         5 &                85\% &                                     I &                                       H/A/O &                \checkmark &                             x &                          x &                          x &                                 x &                              x &                             x &          \checkmark &                          +  \\
\midrule
                                           MovinPy &                             Mobility data analysis &                          P &                         3 &                   0 &                                     I &                                         H &                         x &                             x &                          x &                          x &                    \halfcheckmark &                              x &                             x &                   x &                           - \\
\midrule
                    HuMobi~\cite{smolak2021impact} &                          Human mobility prediction &                          P &                         3 &                   0 &                                     I &                                         H &                \checkmark &                    \checkmark &                          x &                          x &                    \halfcheckmark &                              x &                             x &                   x &                          +  \\
\midrule
                    PTRAIL~\cite{haidri2021ptrail} &             \begin{tabular}[x]{@{}l@{}}Parallelization and \\feature extraction \end{tabular} &                          P &                         4 &                  ** &                                     I &                                       H/A/O &                         x &                             x &                          x &                          x &                                 x &                              x &                             x &          \checkmark &                         ++  \\
\midrule
            TransBigData~\cite{yu2022transbigdata} &                                     Transportation &                          P &                         5 &                90\% &                                    C  &                                         H &                \checkmark &                             x &                          x &                          x &                    \halfcheckmark &                              x &                    \checkmark &          \checkmark &                           - \\
\midrule
           mobilityDB~\cite{zimanyi2020mobilitydb} &                               Storing and querying &               SQL             &                         6 &                97\% &                                     I &                                        H/A &                         x &                             x &                          x &                          x &                                 x &                              x &                             x &      \halfcheckmark &                          +  \\
\midrule
                       Traja~\cite{shenk2021traja} &                                Animal trajectories &                          P &                         6 &                76\% &                                     I &                                         A &            \halfcheckmark &                             x &                          x &                          x &                                 x &                              x &                             x &          \checkmark &                         ++  \\
\midrule
            Tracktable~\cite{wilson2014tracktable} &                             Moving object tracking &                        P/C++ &                         2 &                  ** &                                     I &                                         O &            \halfcheckmark &                             x &                          x &                          x &                                 x &                              x &                             x &          \checkmark &                         ++  \\
\midrule
                                              MEOS &                      Spatio-temporal data analysis &                        C++ &                         4 &                   * &                                    I  &                                       H/A/O &                         x &                             x &                          x &                          x &                                 x &                              x &                             x &                   x &                           + \\
\midrule
                                            MoveTK &     
                                            \begin{tabular}[x]{@{}l@{}}Computational movement\\analysis \end{tabular}
                                              &                        C++ &                         - &                  ** &                                    I/C &                                       H/A/O &                         x &                             x &                          x &                          x &                                 x &                              x &                             x &                   x &                         ++  \\
\midrule
           adehabitatLT~\cite{calenge2011analysis} &                                     Animal habitat &                          R &                         4 &                  ** &                                     I &                                         A &                         x &                             x &                          x &                          x &                                 x &                              x &                             x &                   x &                         ++  \\
\midrule
                                           moveVis &                                      Visualization &                          R &                         6 &                93\% &                                     I &                                         A &                         x &                             x &                          x &                          x &                                 x &                              x &                             x &          \checkmark &                           - \\
\midrule
                stplanr~\cite{lovelace2018stplanr} &     \begin{tabular}[x]{@{}l@{}}Sustainable \\transport planning \end{tabular} &                          R &                         6 &                   * &                                    C  &                                         H &                         x &                             x &                          x &                          x &                                 x &                              x &                             x &                   x &                           - \\
\midrule
                                      trajectories &           
 \begin{tabular}[x]{@{}l@{}}Object tracking \\ and interaction \end{tabular} &                          R &                         5 &                   * &                                     I &                                       O/H/A &                         x &                             x &                          x &                          x &                                 x &                              x &                             x &          \checkmark &                          +  \\
\midrule
                                            TrackR &                           Running and cycling data &                          R &                         6 &                52\% &                                     I &                                         H &                \checkmark &                             x &                          x &                          x &                                 x &                              x &                             x &          \checkmark &                           + \\
\midrule
                                        ArcGIS Pro &                                       Spatial data &                   (P) &                   -        &                  ** &                                    I/C &                                       O/H/A &            \halfcheckmark &                \halfcheckmark &                          x &                          x &                                 x &                              x &                \halfcheckmark &          \checkmark &                         ++  \\
\midrule
\bottomrule
\end{tabular}

}
\label{tab:packages}
\end{table}

Due to the long history of research in transportation, human migration, and animal behavioral research, a large variety of libraries for (human) movement data processing exists. 
\citet{joo_navigating_2020} survey an impressive number of 58 packages for movement analysis in R. Based on this work and the overview provided by~\citet{grasermove}, we selected the libraries that aim at supporting movement analysis in Python, R and C++.
In \autoref{tab:packages}, these selected libraries are compared in terms of their user-friendliness (documentation and robustness), their focus and their provided functionality for human movement data analysis.
To compare packages by the quality of their documentation, we evaluate them on a scale from 0-6 based on criteria used for peer-review of packages by pyOpenSci\footnote{\url{https://www.pyopensci.org/contributing-guide/intro.html}} and ROpenSci\footnote{\url{https://ropensci.org/}}. See appendix \ref{a:documentationscore} for the list of criteria.

Many of the surveyed R libraries have a strong focus on animal behavioral analysis~\citep{joo_navigating_2020} (not all included in \autoref{tab:packages}). The packages that can (also) be applied to human mobility analysis have a focus on basic statistical analysis of trajectories, such as measuring the spatial extent of animal motion (e.g., adehabitatLT~\cite{calenge2006package}), or the duration and distance of movement trajectories (e.g., TrackR~\cite{frick2017tracker}). Currently there is no coherent framework available in R that provides the functionalities required for human movement analysis with a focus on transport applications.
Furthermore, there are several libraries available in C++ such as Tracktable~\citep{wilson2014tracktable}, MEOS\footnote{\url{https://github.com/adonmo/meos}}, and MoveTK\footnote{\url{https://github.com/movetk/movetk}} that promise efficient and fast tools for trajectory data processing, although they may be less accessible for the research audience in human mobility and transportation. Furthermore, these libraries provide only highly specific functionalities and do not represent a comprehensive framework for movement data analysis.

ArcGIS Pro is a proprietary software for general spatial data processing with modules for movement data analysis such as speed and acceleration computation, trackpoint clustering and in particular trajectory visualization\footnote{\url{https://pro.arcgis.com/en/pro-app/2.8/tool-reference/intelligence/an-overview-of-the-movement-analysis-toolset.htm}}. However, the different functionalities are scattered over different toolboxes and ArcGIS Pro does not provide a consistent framework for the analysis of movement data. Due to its proprietary nature, we could not evaluate documentation and testing as we did for the other packages, but we assume both are on a high level. We did not include QGIS\footnote{\url{https://www.qgis.org/en/site/}}, a high quality open-source GIS Project, in the table, as there are no well-maintained plug-ins for movement or trajectory data analysis available. However, QGIS could be used in combination with Python libraries or the mobilityDB~\citep{zimanyi2020mobilitydb} library.

In Python, many open-source libraries have emerged as tools to both facilitate and standardize data processing and analysis. The geographic information science (GIScience) community in particular has benefited significantly from Python libraries, for example, the data models implemented in Shapely~\citep{gillies2013shapely} and the I/O formats for geographic data as offered in the Fiona package\footnote{\url{https://github.com/Toblerity/Fiona}}. Most importantly, spatial data can be handled easily with the Geopandas library~\citep{kelsey_jordahl_2021_5573592} that directly builds up on Pandas~\cite{mckinney2011pandas}, one of the most established Python libraries for data analysis and manipulation.
 
In the past years, Python has become the de-facto standard for data science and machine learning applications, which are increasingly important for the analysis of movement data~\cite{luca_survey_2021, toch_analyzing_2018}. However, only few libraries have attempted to provide preprocessing and analysis tools specifically for human mobility in a comprehensive Python package (see \autoref{tab:packages}). Although many algorithms for trajectory data mining were developed in the last decade~\cite{zheng2015trajectory}, their open-source availability in Python is limited and they often suffer from insufficient documentation and testing standard, such as HuMobi~\citep{smolak2021impact} or MovinPy. Others are well-maintained but limited in scope, such as Traja~\cite{shenk2021traja} that targets animal movement, PTRAIL~\cite{haidri2021ptrail} for parallel processing, and TransBigData~\cite{yu2022transbigdata} which focuses on data analysis on a collective level, similar to the R library stplanr~\cite{lovelace2018stplanr}.

Notable exceptions are MovingPandas~\citep{graser_movingpandas_2019}, scikit-mobility~\citep{pappalardo2019scikitmobility}. MovingPandas is based on Pandas and Geopandas, and focuses on low-level trajectory manipulation such as splitting, merging and visualizing trajectories. On the contrary, the scikit-mobility library~\citep{pappalardo2019scikitmobility} targets high-level analysis functions, including computing human mobility metrics, generating synthetic trajectories and assessing privacy risks. Both libraries are actively maintained and contain various measures to ensure high code quality, but the definition of their data model implies a focus on movement trajectories (MovingPandas) or mobility flows (scikit-mobility), which omits important concepts describing individual human mobility such as activities, trips or tours~\cite{axhausen2007definition}. 

We aim to close this gap with the Trackintel framework that utilizes an established data model from the transportation literature, which incorporates different semantic aggregation levels of tracking data specific to human mobility.

\section{Trackintel framework} \label{sec:trackintel_framework}
\begin{figure}[ht]
    \centering
    \includegraphics[width=\textwidth]{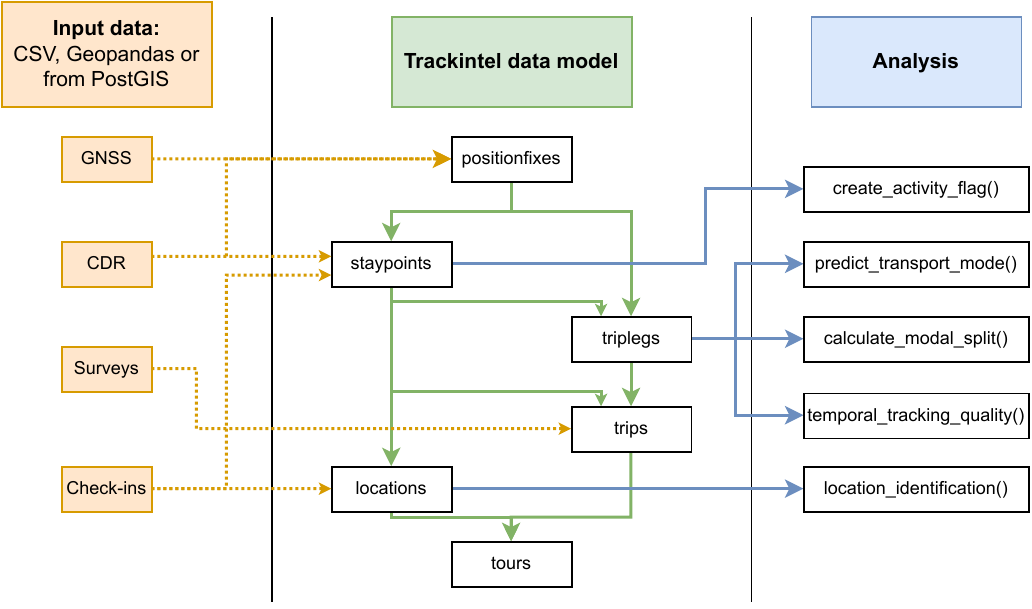}
    \caption{Overview of the Trackintel framework.}
    \label{fig:main}
\end{figure}
Trackintel\footnote{\url{https://github.com/mie-lab/trackintel}} is a library for the analysis of spatio-temporal tracking data with a focus on human mobility. The core of Trackintel is the hierarchical data model for movement data~\citep{axhausen2007definition} that is widely adopted in GIScience~\citep{bucher2019location}, transport planning~\citep{chen2016promises} and related fields~\citep{rout2021using}. We provide easy-to-use and efficient functionalities for the full life-cycle of human mobility data analysis, including import and export of tracking data of various types (e.g, GPS trackpoints, location-based social network (LBSN) check-ins, call detail records), data model generation and preprocessing, analysis, and visualization. A conceptual overview of the different components of Trackintel can be found in \autoref{fig:main}.

Trackintel focuses on the mobility of individual persons or objects (e.g., as opposed to crowd flows) and all functionalities are implemented as user-specific, based on unique user identifiers that link data to the respective tracked users. Trackintel is implemented in Python and is built mainly on top of Pandas~\citep{mckinney2010data} and GeoPandas~\citep{kelsey_jordahl_2021_5573592} using accessor classes, a method to extend Pandas classes\footnote{\url{https://pandas.pydata.org/docs/development/extending.html}}. This design makes Trackintel easy to use for Python users and ensures its broad compatibility with other Python spatial analysis libraries.

\subsection{The Trackintel data model}
\begin{figure}[t]
    \centering
    \includegraphics[width=\textwidth]{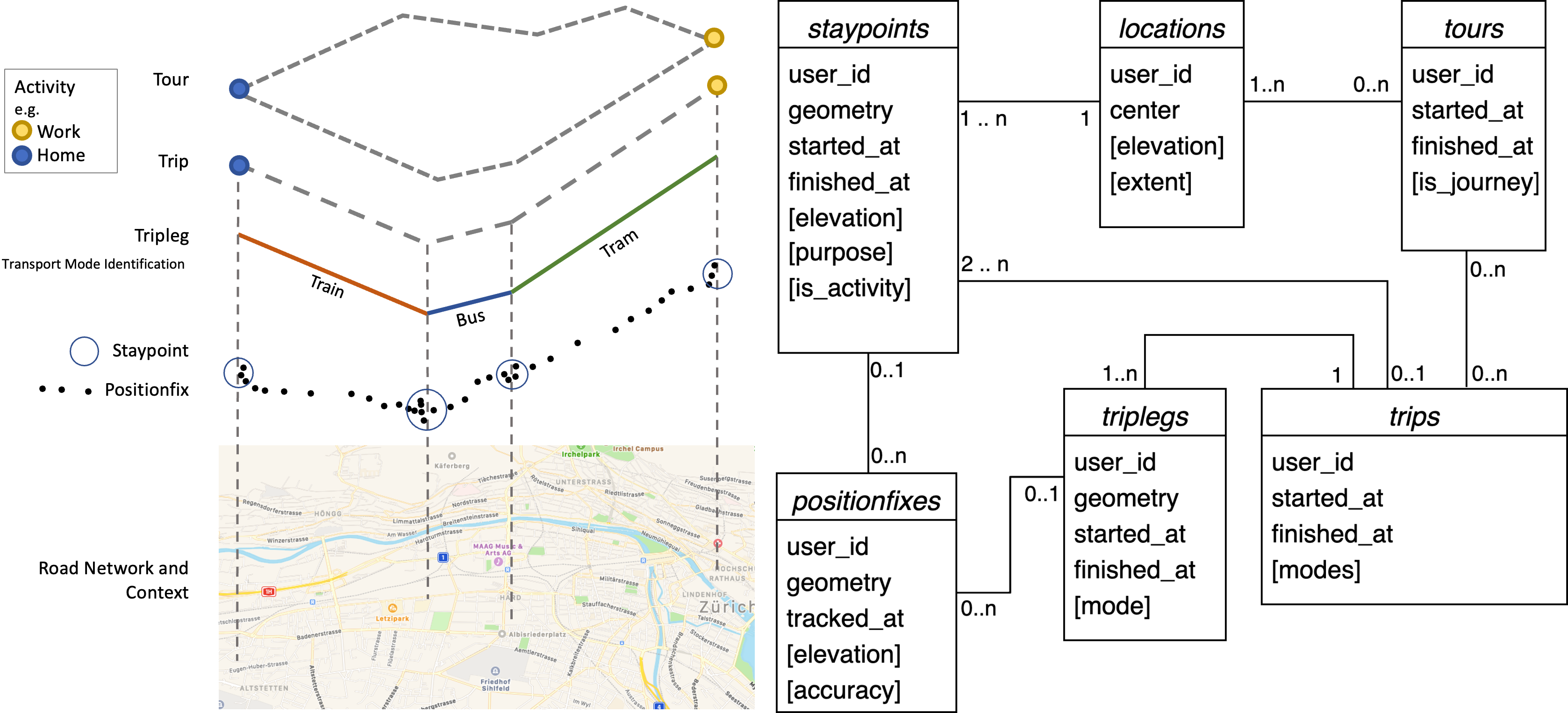}
    \caption{Semantic visualization of the Trackintel data models and their UML diagram, with mandatory and optional attributes (shown in square brackets). The relations between the different classes are shown in the connecting lines. Figure adopted from~\citep{jonietz2018continuous}}
    \label{fig:UML}
\end{figure}
\begin{table}[h!]
\centering
\caption{Description of the mandatory and optional columns for Trackintel data models. }
\resizebox{.9\textwidth}{!}{
\begin{tabular}{@{}llp{9cm}@{}}
\toprule
Data models & Fields                    & Description                                                                                                                          \\ 
\midrule
All          & id                         & The unique identifier for the record                                                                                                 \\
             & user\_id                   & The unique user identifier                                                                                                           \\
             & tracked\_at                & The timestamp for the point (only for positionfix)                                                                                   \\
             & started\_at                & The starting time of the record (except for positionfix and location)                                                                             \\
             & finished\_at               & The ending time of the record (except for positionfix and location)                                                                               \\
Positionfix  & geometries                 & Point geometry                                                                                                                       \\
Staypoint    & geometries                 & Point geometry                                                                                                                       \\
             & purpose (optional)                   & Purpose label for the staypoint. This could be either an activity purpose (e.g., home), or an non-activity   purpose (e.g., wait). \\
             & is\_activity (optional)              & Boolean flag indicating whether the staypoint is an activity                                                                         \\
Location     & center                     & Point geometry representing the center                                                                                               \\
             & extent   (optional)        & Polygon geometry representing the extent                                                                                             \\
Tripleg      & geometries            & Line geometry                                                                                                                 \\
             & mode   (optional)          & Transport mode label                                                                                                                 \\
Trip         & origin\_staypoint\_id      & The identifier of the starting staypoint                                                                                             \\
             & destination\_staypoint\_id & The identifier of the destination staypoint                                                                                          \\
             & primary\_mode   (optional) & The main transport mode label                                                                                                        \\
Tour         & location\_id               & The start and end location identifier                                                                                                \\
             & journey                    & Boolean flag indicating whether the tour is a journey (A tour is called a journey if the start and end location is home).        \\ \bottomrule
\end{tabular}}
\label{tab:modelDescp}

\end{table}
The modeling framework employed by Trackintel is based on the activity-based analysis framework in transport planning, which regards travel demand as derived from our need to perform activities at different locations. We follow the definition from \citep{schonfelder2016urban} that people's daily mobility consists of staying at locations to perform activities and traveling between locations for the next activity~\citep{schonfelder2016urban}. In this definition and following the description in~\citep{axhausen2007definition}, movement is separated from activities at different semantic levels. Trackintel implements six classes to represent movement data in this hierarchical model: \textit{positionfix}, \textit{staypoint}, \textit{tripleg}, \textit{trip}, \textit{tour}, and \textit{location}. \autoref{fig:UML} gives an overview of the hierarchical modeling structure and shows the classes in an UML diagram with their mandatory attributes and optional attributes in square brackets. 
All Trackintel classes are implemented as Pandas Dataframes or Geopandas Geodataframes. In order to be considered a valid Trackintel object, all mandatory attributes have to be present as columns with the correct names as shown in \autoref{fig:UML}. A more detailed explanation of the required and optional attributes of the Trackintel classes is given in \autoref{tab:modelDescp}. Geometries need to be of the defined type, with the exception of the \textit{Location} class that can have multiple geometries. Furthermore, all timestamps for the time fields required by Trackintel have to be timezone-aware\footnote{See \url{https://docs.python.org/3/library/datetime.html\#aware-and-naive-objects} for an explanation}. Besides these formal requirements, classes can contain any additional information required for specific analysis. In the following, the different classes and their semantics are introduced.

\textbf{Positionfix.} 
\textit{Positionfix} is the smallest unit of tracking in the Trackintel data model, which consists of timestamped position records, for example generated by GNSS trackers or call detailed records (CDR) data. Positionfixes are often directly transferred from raw tracking data and are thus a natural entry point to the Trackintel data model, where it can further be processed and segmented into triplegs and staypoints. No inherent semantics are included since movements and activities cannot be distinguished from Positionfix. 

\textbf{Staypoint.}
\textit{Staypoint} represents a point in space, which is defined as an individual remaining within a defined geographical radius for a defined time. Compared to the raw \textit{positionfix} points, staypoints can represent stationary points that carry particular semantics, such as the purpose of the stay, or they can represent an intermediate stay such as waiting for a bus. To distinguish between these two types of staypoints, we introduce the concept of \textit{activity}: an activity staypoint is usually the reason for a person to travel and has an important purpose with attached activity label (e.g., home), while a non-activity staypoint only represents a trivial stationary point (e.g., waiting). The exact definition of an activity depends on the goal of the study. In Trackintel, activities are staypoints with the attribute \textit{activity\_flag} set to \textit{True}, which can be obtained through user labels or directly inferred from data (see section \ref{sec:location_labelling}). While activity \textit{staypoints} are the basic unit for constructing trips, which mark the start and end of a \textit{trip}, non-activity \textit{staypoints} can only be part of a \textit{trip}. Additionally, \textit{staypoints} can be spatially aggregated to form \textit{locations}. 

\textbf{Tripleg.}
The most basic level of movement is defined as \textit{tripleg} (referred to as stage in \cite{axhausen2007definition}), which formally represents a continuous movement without changing transport mode or vehicle. Therefore, triplegs contain semantics about the movement of an individual such as the mode of transport that is stored in the attribute field \textit{mode} if available. This information can be obtained from user labels~\citep{hong2021clustering, zheng2010geolife} or inferred using heuristics directly from the data, which is implemented as labeling functions in Trackintel (see Section \ref{sec:mode_labelling}). Triplegs can be created from positionfixes and can be aggregated to form trips.

\textbf{Trip.}
\textit{Trip} represents all travels between two activities and summarizes all triplegs and non-activity staypoints between two consecutive activity staypoints. Trips inherit the activity purpose from the activity label attribute of the destination staypoint. As they are often the primary quantity of interest in transport planning studies, trips, together with activities, are the core of the movement data model proposed in~\citet{axhausen2007definition}.

\textbf{Location.}
Activity staypoints represent individual visits at places that are significant to the visitor. Trackintel models these significant places using the \textit{location} class to enable the characterization of the place that is visited. While the information attached to staypoints is bound to the individual visit (e.g., the specific activity or the time of day), the semantics of locations are related to the place independent of the visit (e.g., land use or the opening hours of a shop). Locations are modeled with two different geometries, a point geometry for the center of the location and a polygon geometry to describe the extent of a location.

\textbf{Tour.}
The mobility of individuals is centered around a few significant locations that act as the basis of their travel behavior. Individuals conduct several activities and trips if convenient but return home (or to a similar significant location) to plan their next activity. This behavior can be analyzed using the \textit{tours} class, which is defined as ``a sequence of trips starting and ending at the same location''~\cite[p. 4]{axhausen2007definition}, referring to the location class defined above. A special case of a tour is the concept of \textit{journey} that starts and ends at the home location of an individual. In Trackintel, a tour can be flagged as a journey using the \textit{journey} attribute. A tour contains multiple trips, but one trip can also be part of several tours in case they are nested, e.g. the trip from the work location to the supermarket and back is part of a larger journey that started at home.

\subsection{Data model generation}

The core functionality of Trackintel is to generate all classes defined in the movement data model from the raw tracking data. In practice, this refers to the generation of the entire hierarchical movement data model from positionfix data. However, it should be noted that it is not required and often not practical to enter the framework from positionfixes - the framework can be accessed at any semantic level depending on the available data (e.g., location based social network (LSBN) check-ins represent staypoints without the availability of positionfixes; see \autoref{fig:main} for examples of input levels for different tracking data types). The following section presents the implemented preprocessing steps necessary to aggregate data through the hierarchy levels. The output of all \texttt{generate} functions is a (Geo)DataFrame with the fields listed in \autoref{tab:modelDescp}.

\textbf{Generate staypoints.}
In Trackintel, \textit{staypoints} are generated from \textit{positionfixes} using the \texttt{trackintel.preprocessing.positionfixes.generate\_staypoints()} function. It implements an extended version of the sliding window algorithm for staypoint detection first reported in~\citep{li2008mining}. The function accepts predefined distance and time threshold parameters, iterates over all \textit{positionfixes} and determines groups of points that satisfy the thresholds for each tracked individual. Therefore, each output staypoint inherits the starting and ending time from the first and last positionfix that belongs to it, respectively, as well as the mean geometry coordinates of the group of positionfixes. The implemented staypoint detection algorithm extends the algorithm from~\citep{li2008mining} by an option to exclude temporal gaps in the tracking data commonly observed in many datasets due to low temporal coverage. This behavior can be controlled using the \textit{gap\_threshold} parameter which represents the maximum time between two consecutive staypoints so that they are still considered consecutive.

\begin{figure}
    \centering
    \includegraphics[width=0.8\textwidth]{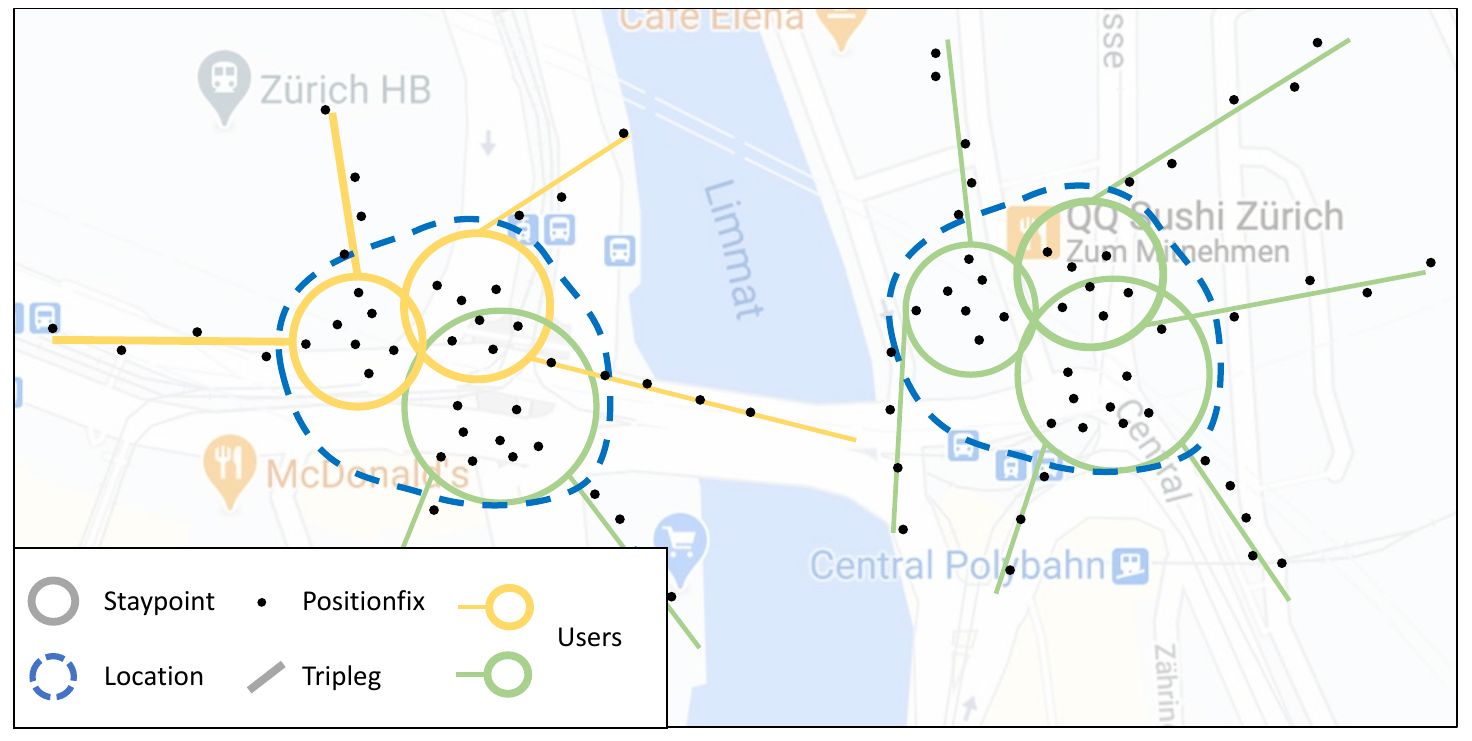}
    \caption{Semantic visualization of the relations between positionfix, staypoint and locations. Staypoints are groups of positionfixes where the users are stationary, and locations are aggregations of staypoints where the user visits multiple times. Locations can be generated across users (left) or for each user individually (right). Map data ©2022 Google.}
    \label{fig:datamodel2}
\end{figure}

\textbf{Generate locations.}
Locations can be generated by aggregating staypoints spatially. We implement the Density-Based Spatial Clustering of Applications with Noise (DBSCAN) algorithm~\citep{jonietz2018continuous, hariharan2004project} in the \texttt{trackintel.preprocessing.staypoints.generate\_locations()} function to aggregate spatially-close staypoints into locations. DBSCAN adopts a set of neighborhood characterization parameters \(\epsilon\) and \(min\_samples\) to depict the tightness of the sample distribution and determine the clustering result. In the context of location generation, \(\epsilon\) controls the distance of which nearby staypoints will be merged into a single location, and \(min\_samples\) determines the minimum number of staypoints to form a location (i.e., how many visits are required at the same place to consider it as significant). Generated locations are equipped with two different geometries. The \textit{center} of the location is a point geometry, calculated as the mean coordinates from all staypoints assigned to the cluster; the \textit{extent} of the location is a polygon geometry, defined as the bounding box of all belonging staypoints. Furthermore, we provide the flexibility in the function to generate locations that are significant to a single user (\autoref{fig:datamodel2} right) or to all users present in a dataset (\autoref{fig:datamodel2} left). While user locations regard staypoints of each tracked user separately in the clustering process and prevent generating locations that are excessively large~\citep{aslak2020infostop}, dataset locations consider all staypoints at the same time and could output locations with shared semantics across users (e.g., train stations or shopping malls). In both options, the center and the extent of the clustered staypoints are attached to the generated locations, providing geometry information that facilitates further processing and analysis tasks. 

\textbf{Generate triplegs.} \label{sec:tpl}
Trackintel implements an algorithm that extracts triplegs from positionfixes based on the assumption that an individual is moving if he or she is not stationary, meaning that all positionfixes that do not belong to any staypoint are assigned to a tripleg. This assignment process is implemented in the \texttt{trackintel.preprocessing.positionfixes.generate\_triplegs()} function, requiring as input the positionfixes with the identifier of the already generated staypoints. Internally, the function aggregates all positionfixes between two consecutive staypoints to form a tripleg, whose line geometry is constructed from connecting the chronically ordered point geometries. Similar to the generation process of staypoints, the start and end timestamp of each tripleg is inherited from the first and last positionfixes that belongs to it, respectively.

\textbf{Generate trips.}
Trackintel implements the method \texttt{trackintel.preprocessing.triplegs.generate\_trips()} to generate trips based on existing staypoints and triplegs. Trips summarize all movement and all non-essential actions (e.g. waiting) between two staypoints that are flagged as activity.
The main result of the trip generation is the identifier management that connects trips with their associated staypoints and triplegs. Trips receive the fields \textit{origin\_staypoint\_id} and \textit{destination\_staypoint\_id} that point to the activities that started and ended the trip, respectively. Furthermore, the function adds the field \textit{trip\_id} to triplegs and non-activity staypoints that occur during a specific trip, and the fields \textit{prev\_trip\_id} and \textit{next\_trip\_id} to activity staypoints that are at the start or the end of a trip. 

The trip detection that is implemented in Trackintel can handle incomplete tracking data and supports the detection of temporal gaps. A temporal gap is defined as missing tracking signals longer than a certain time period~\citep{zhao_applying_2021}, which can be specified using the \textit{gap\_threshold} input parameter to the function. If a temporal gap greater than \textit{gap\_threshold} is detected, we assume the individual performed an unobserved activity and, therefore, the destination of the current and the origin of the next trip is unknown (NaN in the resulting table). 
Finally, the function provides the flexibility to specify whether the trips table should include the geometry using the argument \texttt{add\_geometry}. The geometry of a trip consists of the points for the origin and destination staypoints. If the origin is unknown, we use the first point of the first tripleg instead, or analogously the last point for the destination. 

\begin{figure}
    \centering
    \includegraphics[width=0.8\textwidth]{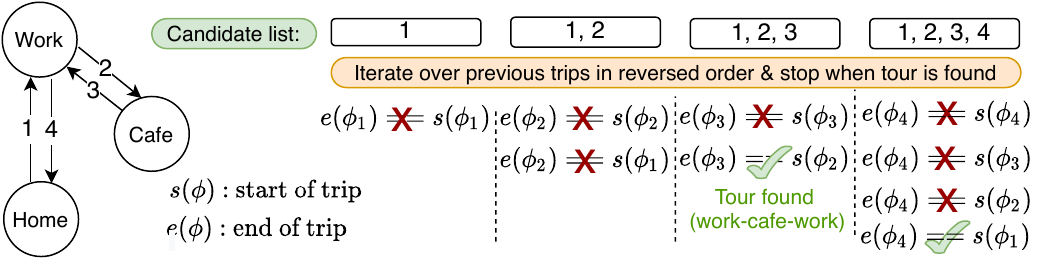}
    \caption{The algorithm of tour generation implemented in Trackintel. A list of start candidates is maintained and iteratively checked for tour-closing trips.}
    \label{fig:tours}
\end{figure}

\textbf{Generate tours.}
To the best of our knowledge, there is no standardized approach yet how to combine trips to tours. Here, we take a rather broad definition of tours that includes nested tours as described in~\cite{axhausen2007definition}, leaving the user the choice to filter the outputs later. An example of a nested tour is shown in \autoref{fig:tours}: the tour Work-Cafe-Work is part of the longer tour Home-Work-Cafe-Work-Home.
This definition implies an n-to-n relationship between trips and tours: One tour contains multiple trips, and one trip can be part of multiple tours. This is reflected in the output of the tour generation algorithm (\texttt{trackintel.preprocessing.trips.generate\_tours()}), where the output table \textit{tours} contains a list of trips as a field, and the table \textit{trips} has a list of tour IDs for each trip.

Our algorithm to generate tours from trips is explained visually in \autoref{fig:tours}. We iterate over the trips that are sorted chronologically, and maintain a list of tour-starting candidates. Each trip $\phi_i$ is a potential candidate to start a tour. At each iteration, that is, for each trip, we first check whether there is a spatial gap between this and the previous trip $\phi_{i-1}$. Two options are implemented: If the table \textit{staypoints} with the attribute \texttt{location\_id} is provided, we compare the location ID of the end of $\phi_{i-1}$ to the one of the start of $\phi_i$, formally $l(e(\phi_{i-1})) = l(s(\phi_{i}))$. Alternatively, if the staypoints are not available, the predefined spatial distance threshold \textit{max\_dist} controls the maximum distance between the end- and start points, i.e. $d\big(e(\phi_{i-1}), s(\phi_{i})\big) \leq max\_dist$. 

Additionally, our implementation offers the possibility to generate tours that are partially observed to accommodate tracking datasets with a low temporal tracking coverage, e.g., mobile phone data-based studies. A parameter \textit{max\_nr\_gaps} determines how many spatial gaps are allowed within a single tour. Note that no gaps are allowed at the start or end of a tour, because a tour must start and end at the same location, or the start- and end-staypoints must lie within the permitted range. 
If the test described above yields a spatial gap between $\phi_{i-1}$ and $\phi_{i}$, and \textit{max\_nr\_gaps}$=0$, the candidate list is reset to $[\phi_{i}]$. Otherwise, a gap is registered.

Next, it is tested whether $\phi_i$ concludes a tour. For this purpose, we iterate over all candidates in the reversed order, such that the shortest possible tour is found first. Thereby we compare the start point of a candidate $\phi_j$ to the end point of $\phi_i$. Again, the points are compared either by the location ID or via the \textit{max\_dist} parameter. If they are the same, the trips $\{\phi_k\ |\ j\leq k \leq i\}$ form a tour, subject to two further conditions: 
A. While iterating over candidates, the encountered gaps are counted and the time duration is checked. The parameter \textit{max\_time} is used to certify whether the tour takes place within an appropriate time period, by default 24 hours. B. When encountering more than \textit{max\_nr\_gaps} in the reversed iteration, or when reaching a candidate that started more than \textit{max\_time} hours ago, the loop ends and no tour is found. \autoref{fig:tours} shows an example where two tours are found after considering $\phi_3$ and $\phi_4$ respectively.  

\subsection{Import and Export} \label{sec:io}
Reading and writing data are important steps in a standard movement data analysis pipeline. To simplify this process, Trackintel provides an I/O module for accessing movement data and storing intermediate or final results to a file or database. Three methods for converting movement data with attached attribute information to Trackintel-compatible formats are provided: Reading from Pandas Dataframes and Geopandas Geodataframes, from csv file formats and from PostgreSQL databases with PostGIS extension. Also, Trackintel implements helper functions to directly load tracking data from publicly available open-source datasets. The following describes these functions in details.

\textbf{Geopandas.}
The recommended way to load new data into the Trackintel framework is via the Pandas/Geopandas reading functions. The Trackintel \texttt{trackintel.io.from\_geopandas} module provides a read function for every Trackintel datatype that accepts GeoDataFrames as input. The functions support renaming and timezone conversion and return a valid Trackintel GeoDataFrame. Through these interfaces with Geopandas, Trackintel can process movement data stored in the most common geospatial file formats (e.g., shapefile, GeoJSON, and GeoPackage).

\textbf{csv file formats.}
The Trackintel \texttt{trackintel.io.file} module can be used to read and write movement data with a csv file. The read functions provide a mapping of the column names to the required Trackintel attributes, and they transform and check the necessary geometry. The write functions provide an easy way to store Trackintel movement data to the disk for caching results or for distributing results.

\textbf{PostGIS.}
Interfaces to access a PostGIS database are defined in Trackintel in the module \texttt{trackintel.io.postgis}, which allows to store or read datasets located in a PostgreSQL database with PostGIS extension using SQL. This enables the use of Trackintel for larger movement data sets. 

\textbf{Dataset readers.} Trackintel additionally provides reading functions for transforming well-known public movement datasets from their raw data representation into the Trackintel data model. As an example, the Trackintel \texttt{trackintel.io.dataset\_reader.read\_geolife()} function reads the raw data from the Geolife dataset \citep{zheng2010geolife} and transforms them to positionfixes. The  \texttt{trackintel.io.dataset\_reader.geolife\_add\_modes\_to\_triplegs()} function adds the transport mode labels to triplegs, which are provided separately for some individuals in the Geolife dataset. The dataset reading functions facilitate and standardize the processing of public movement datasets using Trackintel, which also helps to benchmark new methods on the same data as related work.

\subsection{Pre- and postprocessing}
Trackintel offers several pre- and postprocessing methods such as the simplification of triplegs using the Douglas-Peucker algorithm~\cite{douglas1973algorithms} (the function \texttt{trackintel.preprocessing.triplegs.smoothen\_triplegs()}) or the aggregation of consecutive staypoints (the function \texttt{trackintel.preprocessing.staypoints.merge\_staypoints()}). The later is a common artifact in tracking data of various sources, in which several consecutive staypoints are generated during a single visit to the same location (e.g., due to noise or outliers recordings in GNSS tracking data). The function can aggregate staypoints that are visited close in time. Specifically, we propose to merge two staypoints $s_1, s_2$ of one individual if the following conditions hold: a) $s_1$ and $s_2$ are consecutive in time, b) $s_1$ and $s_2$ are assigned to the same location, c) there is no tripleg registered between $s_1$ and $s_2$, and d) the time gap between the end time point of $s_1$ and the start of $s_2$ is shorter than a predefined threshold $\theta_{max\_time\_gap}$. The input arguments to this method are thus the staypoints (with location id) and triplegs, as well as the threshold $\theta_{max\_time\_gap}$. An optional dictionary can be passed to specify how to aggregate the attributes that cannot be simply accumulated, e.g. the determination of the geometry of the aggregated staypoint. 

\subsection{Analysis} \label{sec:analysis}
While the main functionality of Trackintel is the implementation of the hierarchical data model, the framework also includes advanced analysis functions in order to label transport modes, activity purposes and to assess the tracking quality of each individual.

\subsubsection{Mode labeling}\label{sec:mode_labelling}
Applications in transport planning often require access to the travel modes of an individual~\citep{kim_gps_2022}. Trackintel offers the function \texttt{trackintel.analysis.labeling.predict\_transport\_mode()} to impute the transport mode labels for triplegs. Since Trackintel does not assume the availability of user-provided labels, context or advanced data from the tracking device (e.g., accelerometer), we implement a simple heuristic to determine the travel mode from the tracking data. This classification is done per \textit{tripleg} based on speed. The speed is approximated by the tripleg length (the distance of individual points in its LineString geometry) divided by its total time duration. The triplegs are labeled based on a simple division into slow mobility (<15km/h average speed), motorized mobility (<100km/h) and fast mobility (>100km/h). In future versions, a more in-depth analysis of travel patterns or map matching~\citep{huang2019transport, widhalm2012transport, bachir2018combining, prelipcean2017transportation} 
could be incorporated in Trackintel.

\subsubsection{Location labeling}\label{sec:location_labelling}
An individual's home- and work-locations play a major role for mobility data analysis. As described in Section~\ref{sec:trackintel_framework}, staypoints may be associated with an activity label, but oftentimes this information is not available. We assign ``home'' and ``work'' activity labels to the staypoints with an adapted version of the OSNA algorithm proposed by~\citet{efstathiades2015identification}. The function, provided in \texttt{trackintel.analysis.location\_identification.location\_identifier()}, divides weekdays into rest, work and leisure time frames. The location with the longest accumulated duration in the ``rest'' and ``leisure'' periods is labeled as home, while work is set to the most predominant location in the ``work'' periods. While the original algorithm derives the hours spent at a location from geo-tagged tweets, we take advantage of the \textit{started\_at} and \textit{finished\_at} attributes of a staypoint. Additionally, similar as in the \textit{R} package proposed by \citep{chen_identifying_2021}, we provide a fast method that simply assigns home and work labels to the two locations that are visited more often in the data (in this order). In both cases the locations can optionally be pre-filtered in order to exclude locations with an insufficient number of staypoints or an insufficient length of stay. 

\subsubsection{Modal split}
If mode labels for \textit{triplegs} are available, Trackintel supports the calculation of the modal split of travel. The Trackintel function \texttt{trackintel.analysis.modal\_split.calculate\_modal\_split()} offers three options: Computing the modal split by count (i.e., how many triplegs with this mode exist), by duration (i.e., sum of individual's tripleg duration) or by travelled distance. Furthermore, the frequency can be set according to the Pandas time series frequency syntax\footnote{\url{https://pandas.pydata.org/pandas-docs/stable/user_guide/timeseries.html}}. The Boolean argument \textit{by\_user} determines the aggregation level, i.e. providing the modal split by user or by dataset. An example for one user is visualized in \autoref{fig:modal_split} where the differences between a modal split by count (\autoref{fig:modal_split_count}) and by distance (\autoref{fig:modal_split_distance}) stand out.

\subsubsection{Tracking quality assessment}\label{sec:qualityanalysis}
The function \texttt{trackintel.analysis.tracking\_quality.temporal\_tracking\_quality()} is implemented to assess the temporal tracking quality of a given movement dataset. Temporal tracking quality, here defined as the proportion of time where the user's whereabouts are recorded, is regarded as a basic measure of the temporal resolution of the dataset~\citep{alessandretti_evidence_2018}. The implemented Trackintel function is able to calculate the daily, weekly or overall tracking quality of each user according to the required granularity levels, which enables individual-level temporal resolution assessment, providing support for filtering low-quality users for further analysis. Additionally, tracking quality of hours of the day and weekdays can be obtained for measuring the tracking data quality differences across time periods.

\subsection{Visualization}

\begin{figure}[ht]
    \centering
     \begin{subfigure}[b]{0.32\textwidth}
        \includegraphics[width=\textwidth]{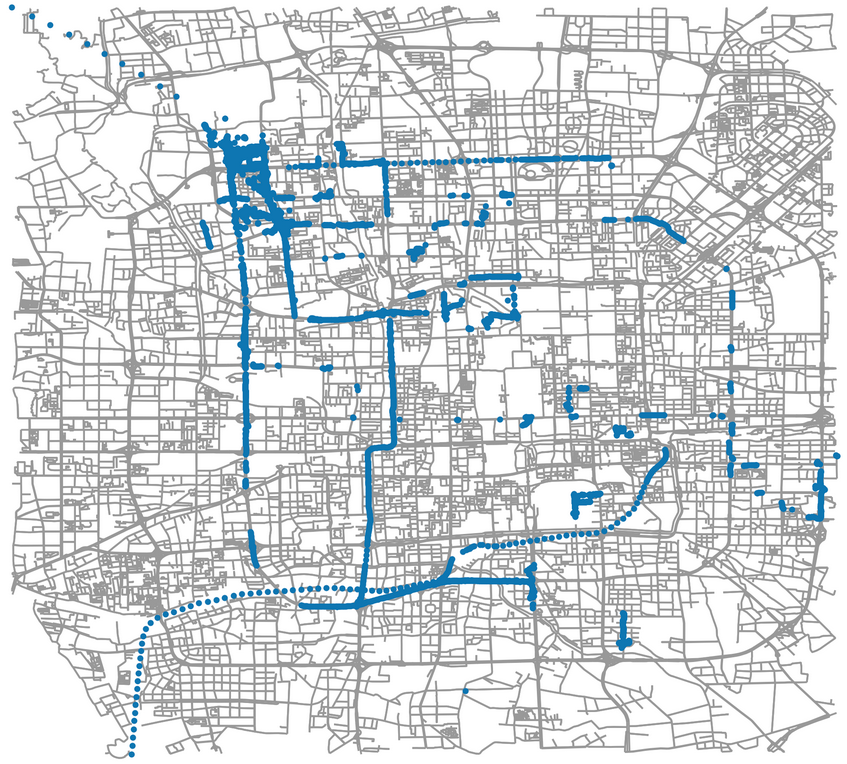}
        \caption{Positionfixes}
        \label{fig:visa}
    \end{subfigure}
    \hfill
    \begin{subfigure}[b]{0.32\textwidth}
        \includegraphics[width=\textwidth]{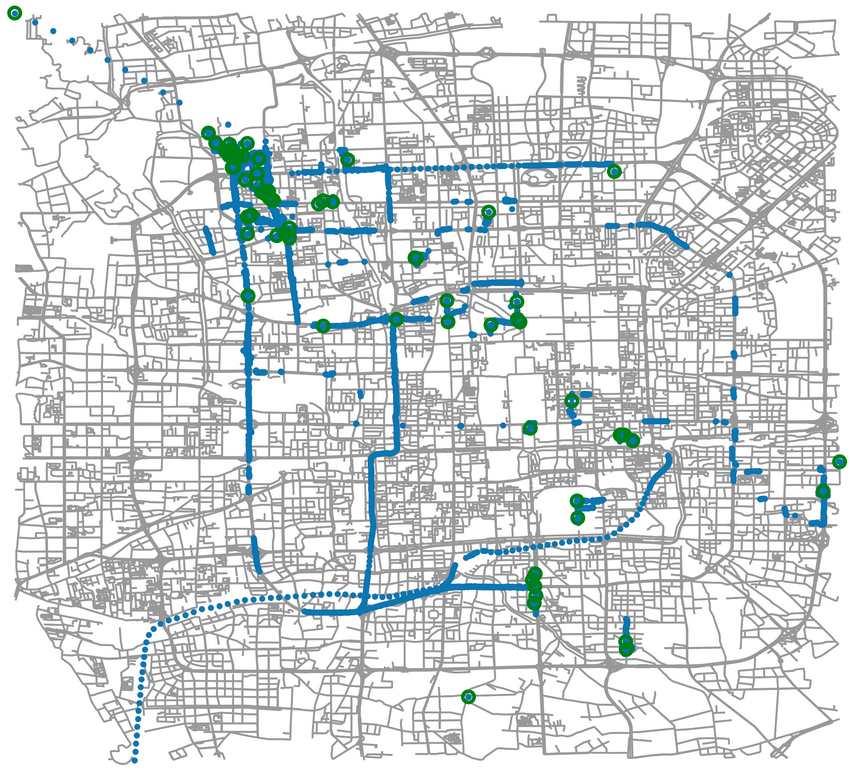}
        \caption{Staypoints}
        \label{fig:visb}
    \end{subfigure}
    \hfill
    \begin{subfigure}[b]{0.32\textwidth}
        \includegraphics[width=\textwidth]{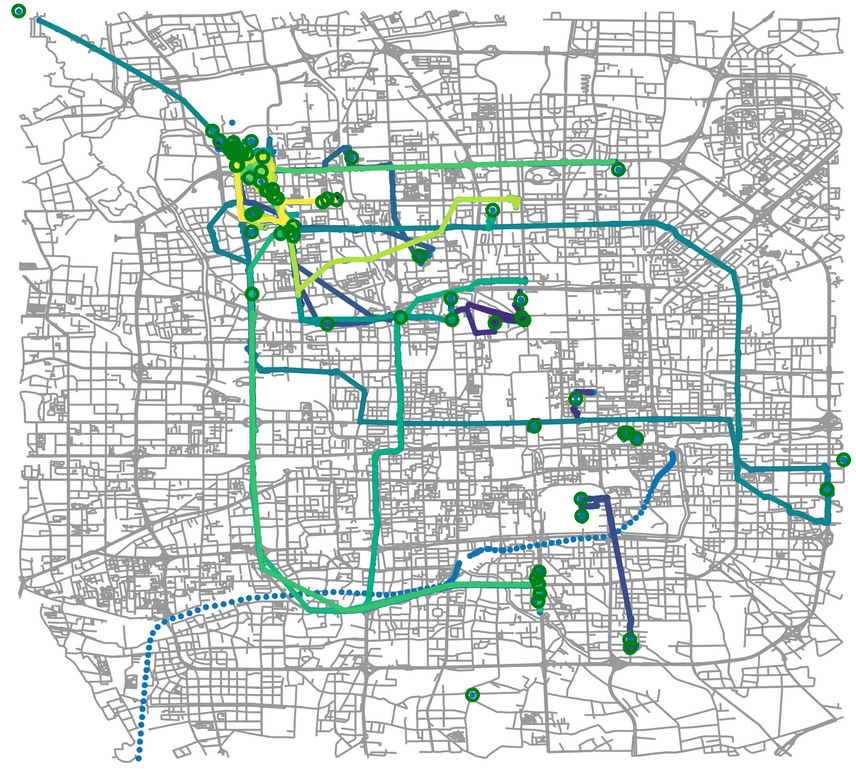}
        \caption{Triplegs}
        \label{fig:visc}
    \end{subfigure}
    \caption{The Trackintel framework offers functions to plot positionfixes (a), staypoints (b), triplegs (c) together with the road network acquired from OpenStreetMaps. This example maps the movements of one  Geolife participant.}
    \label{fig:vis}
\end{figure}

Trackintel provides a \texttt{trackintel.visualization} module that supports the visualization of \textit{positionfixes}, \textit{staypoints} and \textit{triplegs}.
Our implementation standardizes these functions such that each data type can be displayed together with lower aggregation levels (see \autoref{fig:main}). For example, the function \texttt{plot\_locations()} has the optional arguments \textit{positionfixes} and \textit{staypoints}, such that all three can be shown together. In that case, \textit{staypoints} and \textit{locations} are displayed as circles with a predefined radius. 
Furthermore, Trackintel integrates osmnx~\cite{boeing2017osmnx} to optionally show the street network from Open Street Maps as background. \autoref{fig:vis} shows example outputs of the \texttt{plot\_positionfixes()}, \texttt{plot\_staypoints()} and \texttt{plot\_triplegs()} functions for one exemplary participant in the Geolife study.

\begin{figure}
    \centering
    \begin{subfigure}[b]{0.49\textwidth}
    \includegraphics[width=\textwidth]{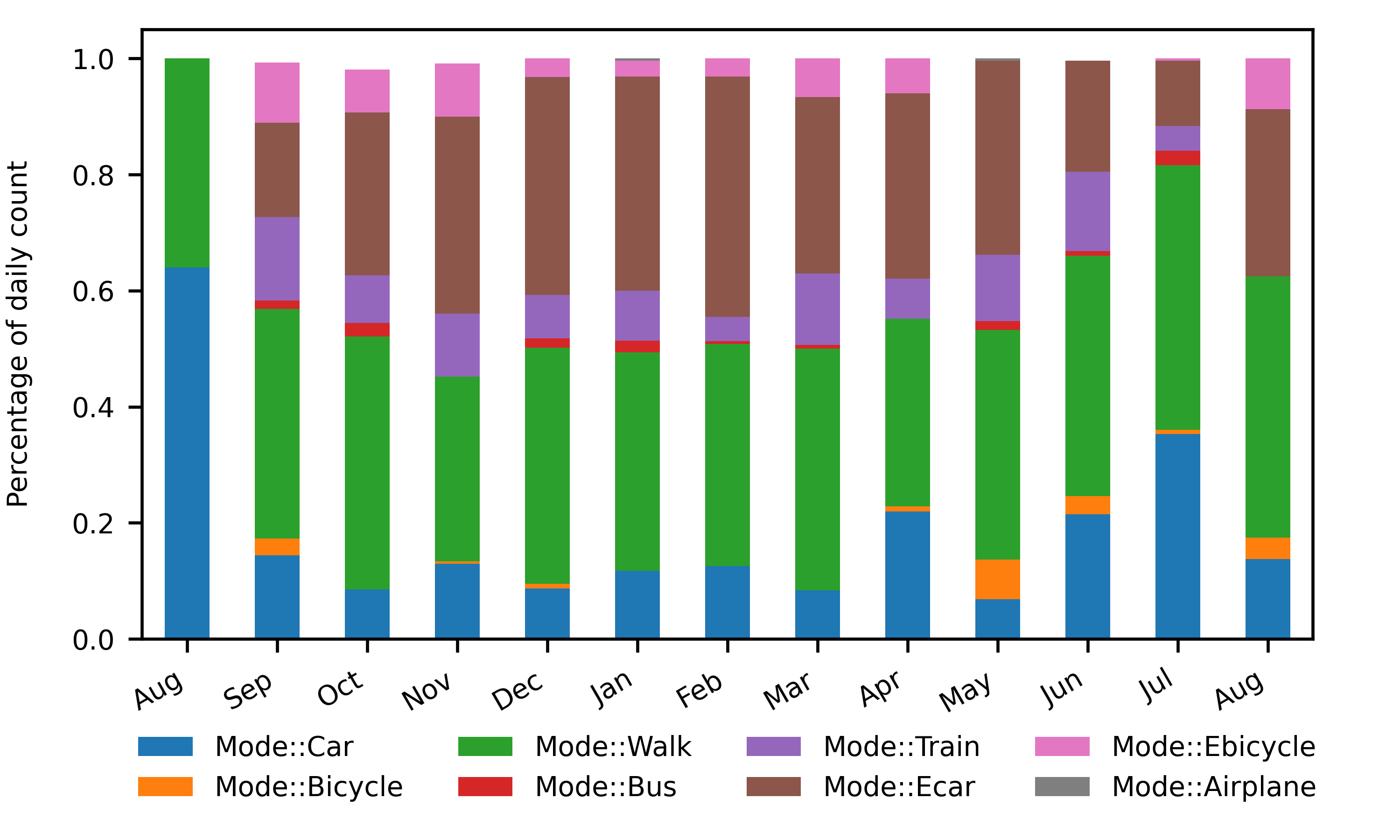}
    \caption{Modal split by count}
    \label{fig:modal_split_count}
    \end{subfigure}
    \begin{subfigure}[b]{0.49\textwidth}
    \includegraphics[width=\textwidth]{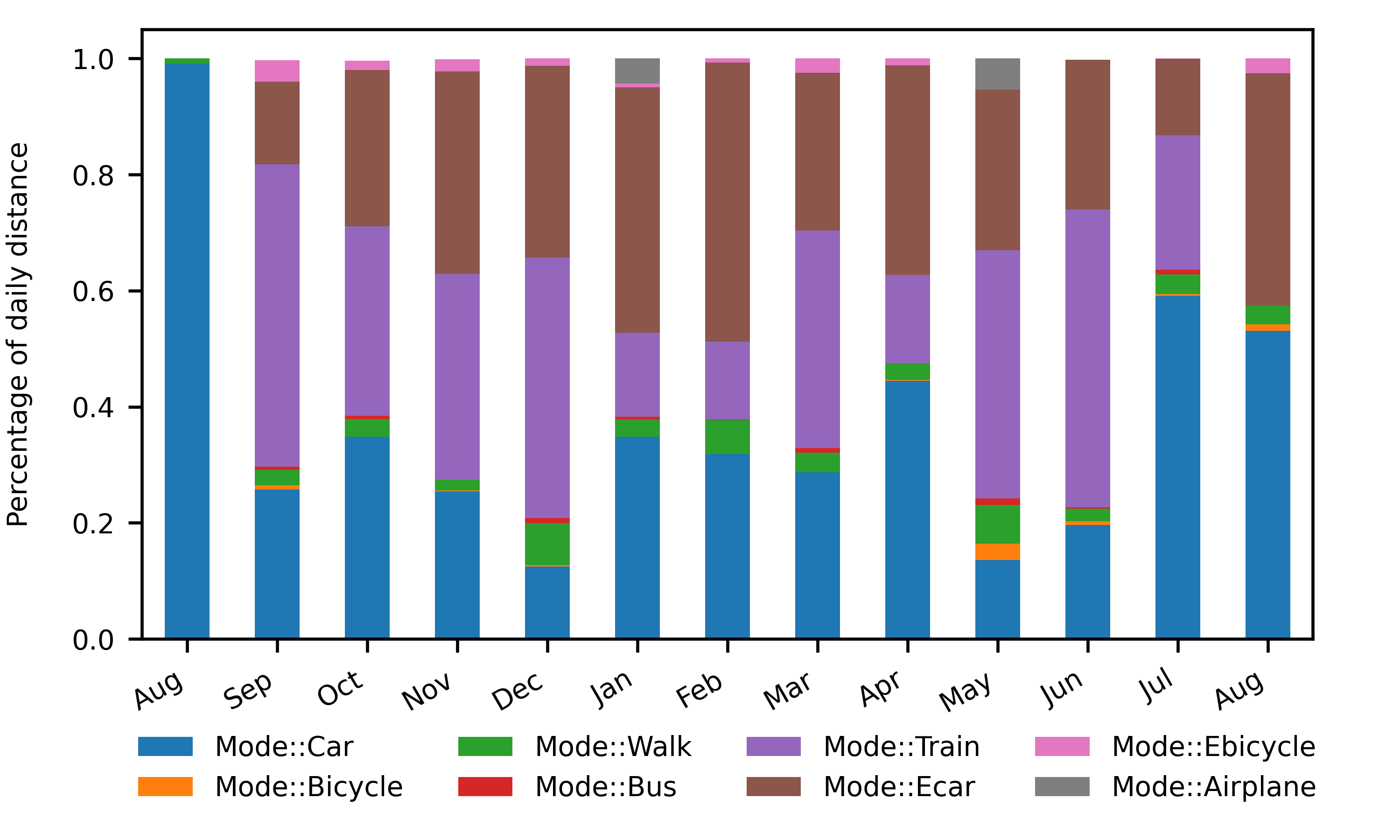}
    \caption{Modal split by distance}
    \label{fig:modal_split_distance}
    \end{subfigure}
    \caption{The visualization result of the Trackintel \texttt{plot\_modal\_split()} function of the triplegs recorded from one Geolife participant. Major differences can be observed between the aggregation by count (number of triplegs) (a) and distance traveled (b).}
    \label{fig:modal_split}
\end{figure}

Finally, Trackintel also provides a flexible method to visualize the development of the modal split over time. The \texttt{plot\_modal\_split()} function shows the output of the \texttt{calculate\_modal\_split()} function from the \textit{analysis} module in a bar plot. Different temporal resolutions (i.e., weeks and months) are handled internally. An example for one user is shown in \autoref{fig:modal_split} where the modal split has been aggregated by month. 

\section{A case study on multiple tracking datasets} \label{sec:case_study}

Trackintel is a framework to standardize mobility data processing and analysis. To demonstrate its capability to handle data from various tracking studies, we provide a case study on four datasets. We read all data from a PostGIS database with the I/O module, preprocess them according to the Trackintel movement data model and compare the datasets in terms of tracking quality, trip characteristics, modal split. The code of the case study is available in the supplemental material and in the public repository\footnotemark.

\footnotetext{\url{https://github.com/mie-lab/trackintel/blob/master/examples/Trackintel_case_study.pdf}}

\subsection{Tracking studies}

We include the data from four tracking studies with two different tracking data types. An overview of the dataset properties is given in \autoref{tab:basicproperties}. The first study is the open-source Geolife dataset~\citep{zheng2009mining} that covers the movement of employees of Microsoft Research Asia, who recorded their movement using GPS trackers. Second, we include two studies that were conducted in collaboration with the Swiss Federal Railway Systems (SBB) under the project name \textit{SBB Green Class}~\citep{martin2019begleitstudie}. In both studies, participants were given full access to all public transport in Switzerland. In addition, the participants from the first Green Class study (Green Class 1) received an electric vehicle and the ones from the second study (Green Class 2) an e-bike. Study participants were tracked with a GNSS-based application (app) called \textit{Myway}\footnote{\url{https://www.sbb.ch/en/timetable/mobile-apps/myway.html}}. The app already provides the data partially preprocessed as staypoints and triplegs. The same app was further used in our fourth dataset, the yumuv study\footnote{\url{https://yumuv.ch/en}} which investigated the impact of a Mobility-as-a-Service app that integrates shared e-scooters, e-bikes and public transport~\cite{martin2021eth}. In the yumuv study, participants were divided into control and treatment group and were tracked for three months.

\begin{table}[ht!]
\caption{Overview of basic features of the considered tracking studies. Locations, staypoints, triplegs, trips and tours are given in multiples of a thousand.}
\resizebox{\textwidth}{!}{
\begin{tabular}{l|rrccrrrrr}
\toprule
{} &  Users &  
\begin{tabular}{@{}r@{}}Tracking period \\ in days (std)\end{tabular}
&                 Input &  Study type &  Locations &  Staypoints &  Triplegs &   Trips &  Tours \\
\midrule
Green Class 1 &    139 &              401 (59) &  Staypoints, Triplegs &   GNSS (app) &      104.5 &       326.9 &     465.2 &  241.8 &   95.0 \\
Green Class 2 &     50 &              314 (76) &  Staypoints, Triplegs &   GNSS (app) &       35.7 &        87.9 &     128.6 &   61.4 &   22.7 \\
Yumuv         &    806 &               87 (38) &  Staypoints, Triplegs &   GNSS (app) &      127.3 &       326.3 &     502.3 &  199.7 &   83.0 \\
Geolife       &    177 &             193 (443) &         Positionfixes &  GPS tracker &       13.6 &        28.9 &      30.2 &   30.2 &    7.2 \\

\bottomrule
\end{tabular}
}
\label{tab:basicproperties}
\end{table}

\subsection{Standardized processing according to the Trackintel data model}

The Trackintel framework offers a straightforward way to transform all data to the same format and to aggregate the data into trips and tours with minimal code. First, the raw GPS data in the Geolife dataset are converted to staypoints and triplegs with the Trackintel \texttt{generate\_staypoints()} and \texttt{generate\_triplegs()} functions. Staypoints are created with a distance threshold of $100$m, i.e. a user must have traveled 100 meters to generate a new staypoint, and a temporal threshold of 30 minutes, as suggested in the original paper~\cite{li2008mining}. Furthermore, consecutive positionfixes with a temporal gap of more than 24 hours in between cannot belong to the same staypoint.

All further preprocessing steps based on staypoints and triplegs are applied with \textit{the same} parameters for all four datasets. This ensures comparability of the results across datasets. More specifically, we derive the user's locations from the staypoints with the \texttt{generate\_locations()} function. The method uses the DBSCAN algorithm with $\epsilon=30$ meters and \(min\_samples = 1\), such that one staypoint is sufficient to form a location. Furthermore, triplegs and staypoints are aggregated to trips with the \texttt{generate\_trips()} function, with input parameter $gap\_threshold = 25$ minutes. At last, tours are generated by merging trips based on a maximum distance (\textit{max\_dist}) of 100m between their start and end points, and with the default parameters \textit{max\_nr\_gaps}=0 and  \textit{max\_time}=24 hours.

\autoref{tab:basicproperties} provides the absolute numbers of locations, staypoints, triplegs, trips and tours per dataset. These quantities decrease from triplegs to trips and tours due to the aggregation steps. Note that for Geolife our parameter choices prevent triplegs from being merged (see \autoref{tab:basicproperties} where the number of triplegs and trips are the same); however, parameters that are more suitable for the trip generation would have decreased the quality of other parts significantly due to the low tracking quality of Geolife. 
In total, the considered datasets include 769,957 staypoints and 1,123,931 triplegs. These quantities depend on the number of participants in the study and the total tracking duration. While the yumuv study has the largest sample size of 806 users, the Green Class 1 study participants have the longest tracking period, with each individual tracked for more than a year on average.

\subsection{Analysis and comparison of tracking datasets}
We now compare the mobility behavior of the study participants of all studies on the trip level as an exemplary usage of the Trackintel \textit{analysis} module. 
The insights from this analysis are summarized in \autoref{tab:data_analysis}.
First, we can derive the number of daily trips per individual from the absolute numbers given above. The study participants in Green Class 1 and Green Class 2 are most active in conducting trips. The low number of trips for Geolife users may be due to low temporal tracking coverage of the dataset. 
Furthermore, we compare the average trip distances and duration across datasets. Interestingly, yumuv and Geolife users take longer trips on average in terms of duration. There is also a clear effect of the bias of yumuv participants towards urban areas, where the trips cover much shorter distances. The number of trips per tour and the number of triplegs that are part of the same trip do not differ much between studies.

\begin{table}[ht]
\caption{Overview of the mobility statistics for the considered tracking datasets.}
\resizebox{\textwidth}{!}{
\begin{tabular}{lrrrrrr}
\toprule
{} &  Trips per day &  Trips per tour &  Legs per trip & Trip distance in km (std) & Trip duration (std) & Tracking quality (std) \\
\midrule
Green Class 1 &           4.32 &            2.73 &           1.92 &        27.4 (478.7) &         0.52 (0.73) &            0.85 (0.17) \\
Green Class 2 &           3.80 &            2.66 &           2.09 &        33.7 (568.2) &         0.51 (0.75) &            0.75 (0.24) \\
Yumuv         &           3.13 &            2.11 &           2.51 &        16.9 (100.4) &         0.68 (0.91) &            0.77 (0.23) \\
Geolife       &           1.70 &            2.37 &           1.00 &       36.1 (3163.5) &         0.64 (0.94) &             0.4 (0.32) \\
\bottomrule
\end{tabular}
}
\label{tab:data_analysis}
\end{table}

\begin{figure}[!htbp]
    \centering
    \includegraphics[width=0.7\textwidth]{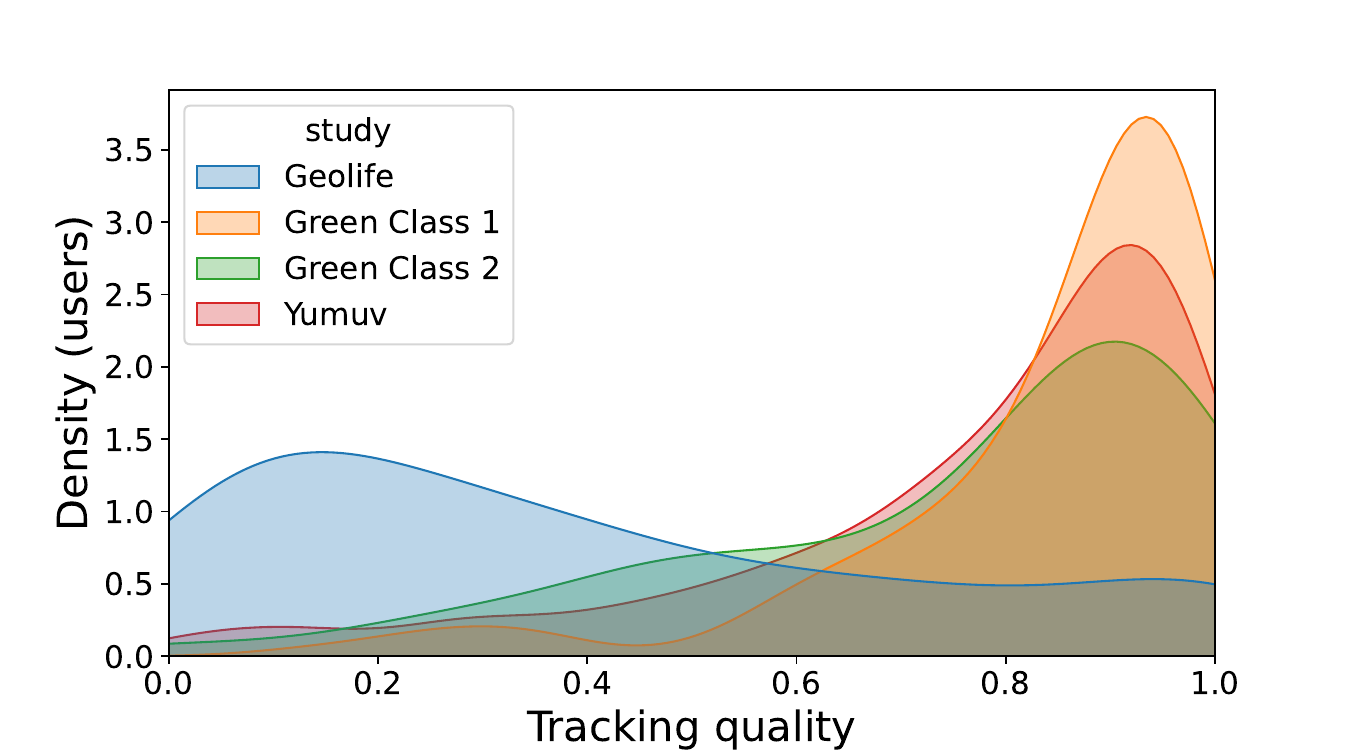}
    \caption{Distribution of the individual temporal tracking quality for the considered datasets. 
    }
    \label{fig:tracking_quality_user}
\end{figure}

Another key part of tracking data analysis regards the temporal tracking quality of a dataset. Here, temporal tracking quality is defined as the temporal coverage of the tracking data (i.e., the completeness) and is computed with the Trackintel function \texttt{temporal\_tracking\_quality()} as explained in section~\ref{sec:qualityanalysis}. The results are given in the last column of \autoref{tab:data_analysis}. The three GNSS-based studies show a high coverage of more than 75\% on average per user, whereas Geolife data only covers about 40\% of the time on average per user. \autoref{fig:tracking_quality_user} shows the distribution of the tracking quality over users. In the Geolife dataset, the temporal tracking quality largely differs across individuals. In comparison, the large majority of Green Class 1 participants reached a coverage of more then 0.7. The large difference between Geolife and the other datasets can be explained by the different hardware that was used in the studies. While the Geolife individuals were equipped with dedicated GPS-only trackers that are prone to localization problems when indoors or in urban canyons, the participants in the Green Class and yumuv studies were tracked with an app on their smart phone that uses the location API of the operating system. The latter has access to all GNSS systems in addition to GPS and can fall back to other technologies such as WIFI or cell tower triangulation if no satellite is available.

\begin{figure}[!htbp]
    \centering
    \begin{subfigure}[b]{0.49\textwidth}
        \includegraphics[width=\textwidth]{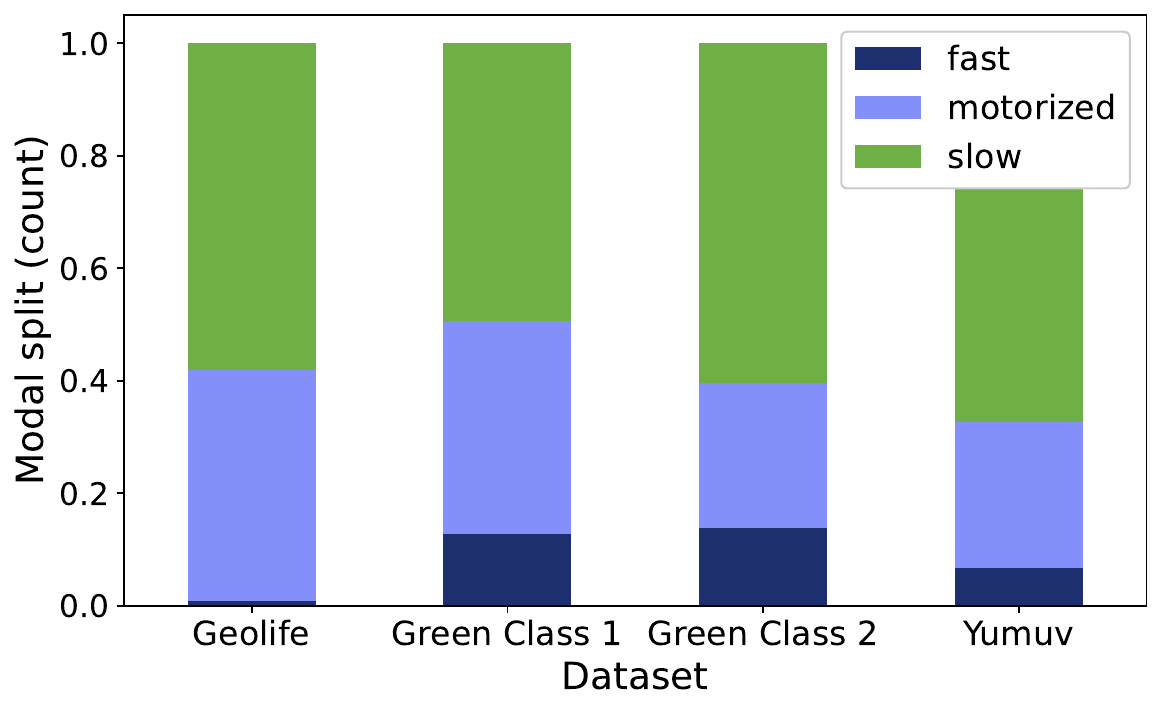}
        \caption{Split by count}
    \end{subfigure}
    \hfill
    \begin{subfigure}[b]{0.49\textwidth}
        \includegraphics[width=\textwidth]{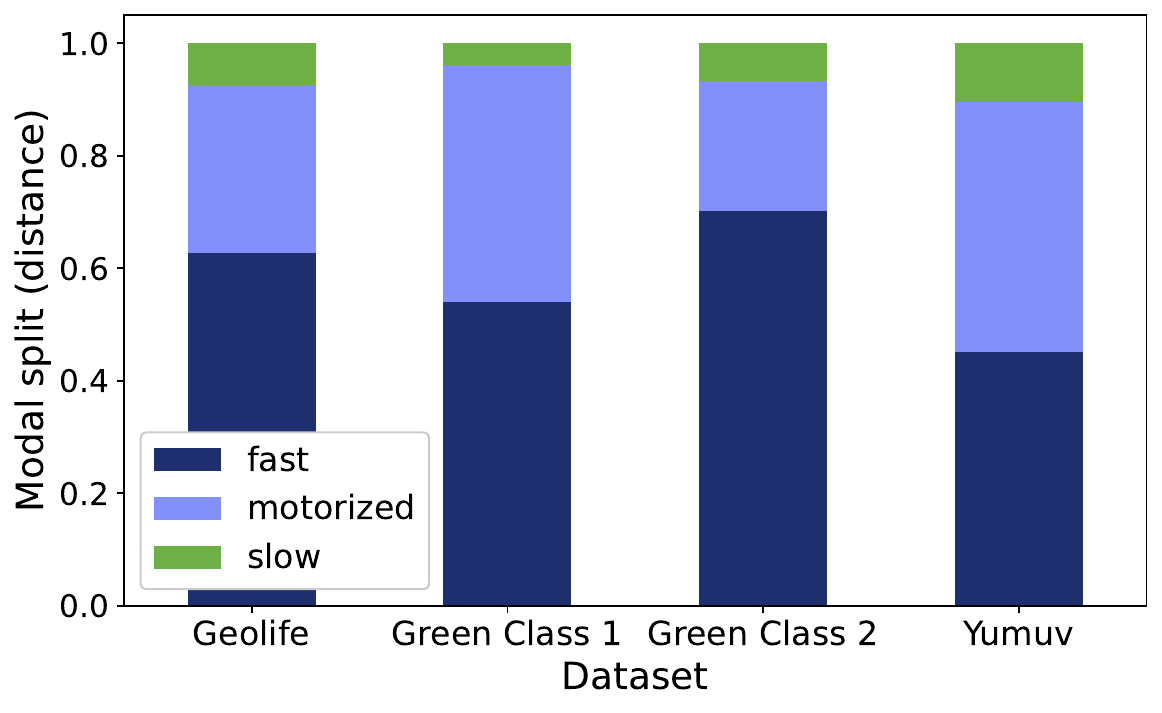}
        \caption{Split by distance}
    \end{subfigure}
    \caption{Comparison of modal split between datasets. The users of different studies differ considerably in terms of their usage of slow, motorized or fast transport.}
    \label{fig:modal_split_studies}
\end{figure}

We further compare the modal split of the tracking studies. The split is computed first as the number of triplegs per mode and secondly as the covered distance per mode. 
We use the Trackintel function \texttt{predict\_transport\_mode()} to approximate the modes for the Geolife dataset, since the original mode labels are not available for all participants and not all the time.
In all other studies, high-quality mode labels are provided, and we aggregate them into the simplified categories of slow mobility (walk, bicycle, scooter), motorized mobility (tram, bus, car and motorbike) and fast mobility (airplane and train). The results are shown in \autoref{fig:modal_split_studies}. The datasets differ significantly with respect to their modal split, which can be explained by the study target group, for example, Green Class participants were given full access to all public transport in Switzerland and are thus more likely to use trains (fast transport). Yumuv individuals on the other hand mostly live in urban areas and they were using the yumuv bundle of shared e- bicycles and scooters, which explains the higher proportion of slow mobility for yumuv. 

\begin{figure}[!htbp]
    \centering
    \begin{subfigure}[b]{0.49\textwidth}
        \includegraphics[width=\textwidth]{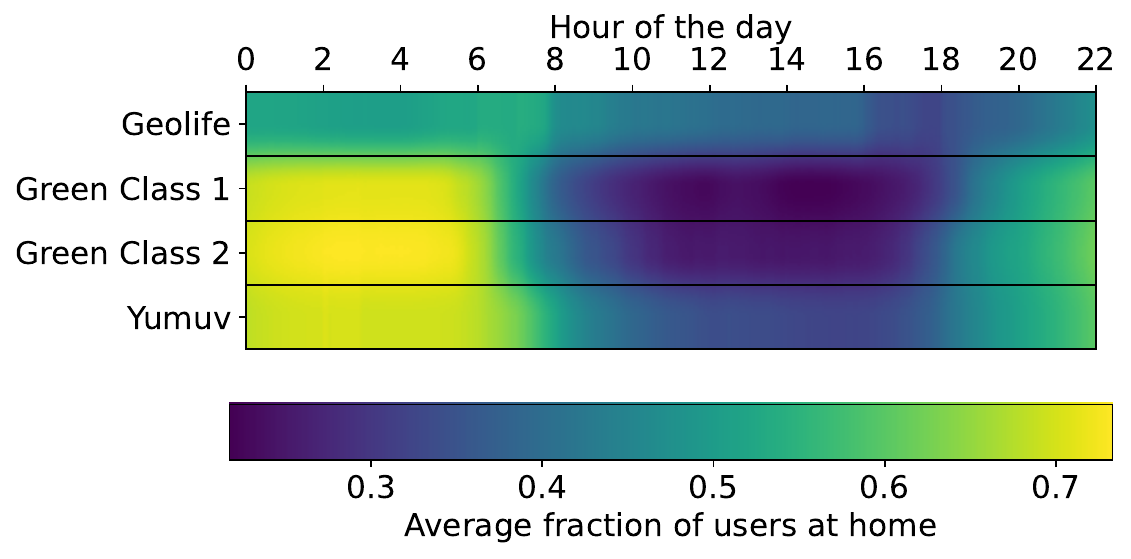}
        \caption{Home-labeled staypoints}
        \label{fig:home}
    \end{subfigure}
    \hfill
    \begin{subfigure}[b]{0.49\textwidth}
        \includegraphics[width=\textwidth]{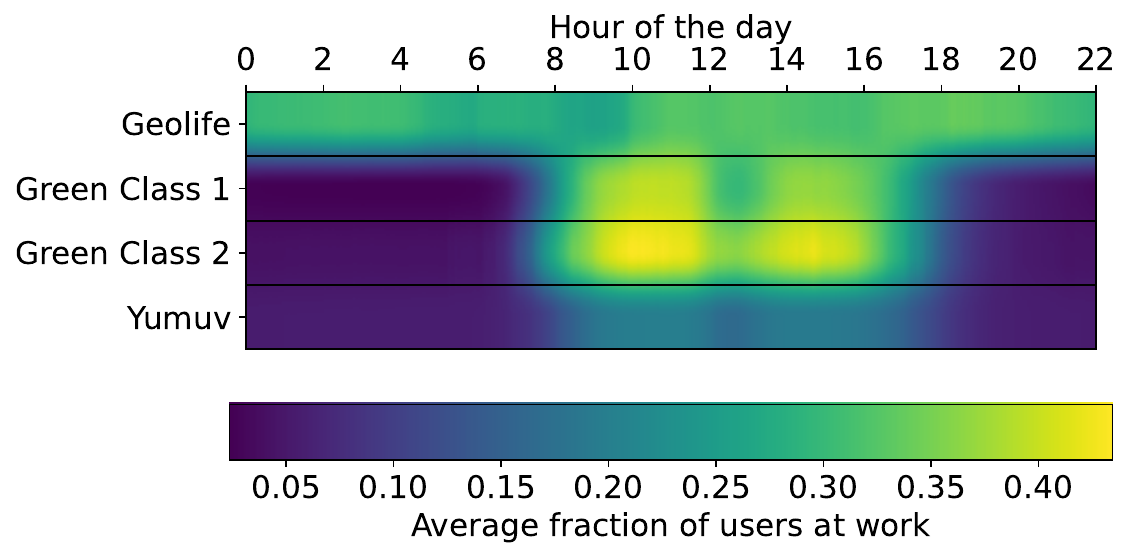}
        \caption{Work-labeled staypoints}
        \label{fig:work}
    \end{subfigure}
    \caption{Distribution of activities over time.}
    \label{fig:labelling}
\end{figure}

Finally, we analyze the daily activity patterns of individuals. Specifically, the time periods where the individuals are at home and at work are computed. For the Green Class 1 \& 2 studies, the activity-label for each staypoint is provided by the participants. For the Geolife and yumuv datasets, on the other hand, we adopt the Trackintel \texttt{location\_identifier()} function that implements the OSNA algorithm~\citep{efstathiades2015identification} to infer the home and work locations. In \autoref{fig:labelling}, the distribution of home and work staypoints over the course of a day is shown. Specifically, the average fraction of users with a staypoint labeled home (or work respectively) is shown for every minute of the day. The fraction of users at home (work) is thereby computed as the number of staypoints per day divided by the number of actively tracked users, where a user is actively tracked if there is at least one staypoint on that day.
The working time between 8am and 5pm as well as the lunch breaks are clearly visible in \autoref{fig:work} for Green Class 1 \& 2 and yumuv, although there are fewer work-staypoints for yumuv. While the home location is reliably identified for both yumuv and Geolife, the identification of the work location seems impaired for the Geolife dataset. As the OSNA algorithm simply selects the second-most visited location as work if the ``home'' and ``work'' labels overlap, the low tracking quality of the Geolife dataset (see \autoref{fig:tracking_quality_user}) could have affected the accuracy of the identification.

In summary, our study demonstrates the ease of comparing data from different sources on all levels of the movement data model and concerning various labels for the movement data. The standardized preprocessing functions implemented in Trackintel also help compare methods and explain possible discrepancies in the analysis results from the different datasets.

\section{Discussion and Conclusion} \label{sec:conclusion}
Quantitative analysis of human mobility currently suffers from a lack of a common data model for movement data and a standardization of preprocessing steps, limiting the  reproducibility and comparability of scientific studies. This article presented Trackintel, a new open-source tool to address these problems. Trackintel implements a widely accepted conceptual data model for movement data and provides functionalities for the full life-cycle of human mobility data analysis: import and export of tracking data collected through various methods, preprocessing, data quality assessment, semantic enrichment, quantitative analysis and mining tasks, and visualization of data and results.

A particular strength of Trackintel is that it greatly simplifies the joint analysis of several movement datasets with different properties. This was shown in a case study where four different datasets where jointly preprocessed and analyzed. We used the analysis methods implemented in Trackintel to compare the datasets with respect to their trip properties, their tracking quality, their modal split and their daily activity patterns. It was demonstrated that rich insights about the characteristics of different tracking datasets can be easily obtained in Trackintel with few lines of code.

On the other hand, the compatibility of Trackintel with diverse datasets limits 
the capabilities of the analysis model. 
A good example is the transport mode prediction function provided by Trackintel that is based on a simple heuristic. A more sophisticated and powerful method can in principle be implemented for a specific dataset, however, the applicability of this method to other datasets will be limited by the availability of specific input data or additional context data.

Importantly, the purpose of Trackintel is not to provide a comprehensive set of analysis functions, but rather a high-quality implementation of standard aggregation and semantics-enrichment steps that are relevant for most tracking studies. 
Nevertheless, Trackintel will be continuously extended to incorporate the latest processing and analysis algorithms and to offer a wider variety of options for the preprocessing, analysis and visualization of movement data.

Finally, Trackintel does not aim to cover all preprocessing and analysis needs for every movement data study. However, due to the compatibility with Pandas and Geopandas, Trackintel can easily be integrated in a larger workflow that comprises a variety of Python data and spatial analysis libraries. In particular, it is targeted at providing the same reliability as these standard libraries. This is achieved through a strong compliance with Python library standards, including a high coverage of unit tests with both real and synthetic data, code reviews as quality checks and continuous integration pipelines. In this setup, new algorithms can be contributed easily without risking to break existing functionality. We therefore believe Trackintel can serve as a standard and well-trusted mobility processing tool. 

\section{Acknowledgement}
This work was supported by the Swiss Data Science Center [C17-14] and the ETH Zurich Foundation [MI-01-19];
Additionally we would like to thank Christof Leutenegger, Sven Ruf, and Nishant Kumar for their code contributions to Trackintel, David Jonietz for helping to create the idea of Trackintel, and René Buffat and Jiří Kunčar for their technical input in the early stage of this project.

\bibliographystyle{plainnat}
\bibliography{references}
\newpage
\appendix
\section{Documentation score} \label{a:documentationscore}
\subsection{Python}
The documentation score reported in \autoref{tab:packages} for python libraries is based on the pyOpenSci package peer-review evaluation critera\footnote{\url{https://www.pyopensci.org/contributing-guide/intro.html}}

\begin{itemize}
    \item Has an Open Software Initiative (OSI) approved license.
    \item Contains a README with instructions for installing the development version.
    \item Contains a vignette (notebook) with examples of its essential functions and uses.
    \item Has a test suite.
    \item Has continuous integration, such as Travis CI, AppVeyor, CircleCI, and/or others.
    \item Includes documentation with examples for all functions.
\end{itemize}

\subsection{R}
The documentation score reported in \autoref{tab:packages} for R libraries is based on the ROpenScie package peer-review evaluation critera\footnote{\url{https://ropensci.org/}}
\begin{itemize}
    \item Does the package have a CRAN accepted license?
    \item The package contains a reasonably complete readme with devtools install instructions.
    \item The package contains a vignette with examples of its essential functions.
    \item The package contains unit tests.
    \item The repository has continuous integration with Travis and/or another service.
    \item Package available on CRAN?
\end{itemize}
    
\end{document}